\documentclass[aps,pra,twocolumn,showpacs,floatfix,superscriptaddress,10pt]{revtex4-1}

  \usepackage[utf8]{inputenc}
  \usepackage[T1]{fontenc}
  \usepackage[margin=2cm]{geometry}
  \usepackage{float}
  \usepackage{color}
\usepackage{graphicx}
\usepackage{amsmath}
\usepackage{amssymb}
\usepackage{bbm}
\usepackage{indentfirst}
\usepackage{dcolumn}
\usepackage{soul}
\usepackage{mathtools}

\usepackage{subfigure}
\usepackage{multirow}
\usepackage{setspace}

\DeclareMathOperator{\Tr}{Tr}
\newcommand{\unn}[1]{ \underline{\underline{#1}} }    
\newcommand{\mxrd}[1]{\unn{#1}}                       
\newcommand{\vrd}{\vec}                               

\title{}

\begin{document}

\title{Magnetic structure of monatomic Fe chains on Re(0001): emergence of chiral multi-spin interactions}
\author{A. L\'aszl\'offy}
\email{laszloffy@phy.bme.hu}
\affiliation{Department of Theoretical Physics, Budapest University of Technology and Economics, Budafoki \'{u}t 8., HU-1111 Budapest, Hungary}
\author{L. R\'ozsa}
\affiliation{Department of Physics, University of Hamburg, D-20355 Hamburg, Germany}
\author{K. Palot\'as}
\affiliation{Department of Theoretical Physics, Budapest University of Technology and Economics, Budafoki \'{u}t 8., HU-1111 Budapest, Hungary}
\affiliation{{Institute for Solid State Physics and Optics, Wigner Research Center for Physics, Hungarian Academy of Sciences, P.\ O.\ Box 49, H-1525 Budapest, Hungary}}
\affiliation{MTA-SZTE Reaction Kinetics and Surface Chemistry Research Group, University of Szeged, H-6720 Szeged, Hungary}
\author{L. Udvardi}
\affiliation{Department of Theoretical Physics, Budapest University of Technology and Economics, Budafoki \'{u}t 8., HU-1111 Budapest, Hungary}
\affiliation{MTA-BME Condensed Matter Research Group, Budapest University of Technology and Economics, Budafoki \'{u}t 8., HU-1111 Budapest, Hungary} 
\author{L. Szunyogh}
\affiliation{Department of Theoretical Physics, Budapest University of Technology and Economics, Budafoki \'{u}t 8., HU-1111 Budapest, Hungary}
\affiliation{MTA-BME Condensed Matter Research Group, Budapest University of Technology and Economics, Budafoki \'{u}t 8., HU-1111 Budapest, Hungary} 

\date{\today}
\pacs{} 

\begin{abstract}
We present results of first-principles calculations of the magnetic properties of Fe chains deposited on the Re(0001) surface. By increasing the length of the chain, a transition is found from an almost collinear antiferromagnetic state for a five-atom-long chain to a spin spiral state with the rotational plane slightly tilted from the surface of the substrate for the 15-atom-long chain. It is shown that a classical spin model derived from the \textit{ab initio} calculations containing only two-spin interactions supports opposite chirality of the spin spiral compared to a direct optimization of the spin configuration within the \textit{ab initio} method. The differences between the results of the two methods can be understood by introducing chiral four-spin interactions in the spin model.
\end{abstract}

\maketitle
\section{Introduction}

The investigation of clusters of magnetic atoms on nonmagnetic surfaces has recently opened several intriguing prospects for the storage and transfer of information on the nanometer scale. The reduced dimensionality of the clusters often leads to an enhancement of the magnetic anisotropy energy \cite{Gambardella}, stabilizing the magnetic structure in one of two states connected by time-reversal symmetry. These two states can in turn be used for designing logic gates \cite{Khajetoorians}. Besides the anisotropy, the interactions between the magnetic adatoms mediated by the substrate also crucially influence the magnetic state. One prominent type of these couplings is the Dzyaloshinskii--Moriya (DM) interaction \cite{Dzyaloshinsky1958,Moriya1960}, the presence of which can be attributed to the spin--orbit coupling and the inversion-symmetry breaking caused by the surface. This interaction leads to the formation of noncollinear structures with a preferred chirality by which the information may be encoded \cite{Menzel}. Linear chains of magnetic atoms on a superconducting surface also offer a possibility for realizing Majorana bound states \cite{Alicea} as fundamental elements of topological quantum computing, the signatures of which have been investigated experimentally in chains both with collinear \cite{Nadj-Perge} and noncollinear \cite{Kim2018} magnetic ground states.

Determining the ground state for an interacting magnetic system based on \textit{ab initio} electronic structure calculations remains a considerable challenge. Such computations can efficiently be performed by mapping the energy or grand potential of the system to a classical spin model. It was demonstrated in Ref.~\cite{Liechtenstein1987} how Heisenberg exchange interactions between pairs of spins may be determined based on the derivatives of the energy, that is, the torque acting on the spin directions. The torque method has been generalized to tensorial 
two-spin interactions appearing in the presence of spin--orbit coupling \cite{Udvardi2003,Ebert2009}, and it was validated for various systems over the last decade \cite{Antal2008,Vida2016,Simon2018}. A fully real-space calculation of the interactions in a magnetic cluster 
is presented in Ref.~\cite{Cardias2016}. However, only considering two-spin interactions in the spin model is not sufficient for describing all types of magnetic order. It was demonstrated in various ultrathin film systems that isotropic four-spin interactions may stabilize up-up-down-down states \cite{Al-Zubi,Kronlein,Romming}, conical spin spirals \cite{Yoshida2012,Zimmermann2014}, or nanoskyrmion lattices \cite{Heinze}.

The problem of finding a spin model which contains all types of magnetic interactions relevant in the system may be circumvented by updating 
the directions of the magnetic moments during the \textit{ab initio} calculations. Because of the higher number of degrees of freedom the computational complexity increases dramatically, but such methods enable a more accurate determination of the magnetic ground state. It was proposed in Refs.~\cite{Stocks,Ujfalussy} that the constrained local moment method within density functional theory is applicable for performing first-principles spin dynamics simulations. Using this method, it was demonstrated in Ref.~\cite{Ujfalussy2004} that the reduction of the symmetry leads to a canted magnetic configuration in a finite Co chain along a step edge on the Pt(111) surface. An alternative procedure for updating the spin directions based on the Landau--Lifshitz--Gilbert equation \cite{Landau,Gilbert} was introduced in Ref.~\cite{Rozsa2014}, where the torques acting on the spins are determined directly from the electronic structure at each time step within a fixed electronic potential.

In the present paper the magnetic properties of monatomic Fe chains are investigated on the Re(0001) substrate. Spin-polarized scanning tunneling microscopy measurements performed for a 40-atom-long Fe chain on superconducting Re in Ref.~\cite{Kim2018} revealed a spin spiral ground state with both in-plane and out-of-plane spin components and a period of approximately four lattice constants. 

The paper is organized as follows. In Secs.~\ref{sec2a} and \ref{sec2b} the details of the \textit{ab initio} calculations are discussed, performed using the Vienna \textit{Ab-initio} Simulation Package (\textsc{vasp}) \cite{VASP} and the embedding technique within the KKR method \cite{Lazarovits2002}, respectively. 
In Sec.~\ref{sec2c} the spin model including two-spin interactions is introduced. The results for the parameters entering the spin model are discussed in Sec.~\ref{sec3a}. The possible magnetic ground states of the chains obtained from the spin model and from a direct optimization within the \textit{ab initio} method are compared in Sec.~\ref{sec3b}. The deviations between the different methods observed for the chirality of the spin structure for the 15-atom-long chain are resolved by taking into account four-spin chiral interactions introduced in Sec.~\ref{sec3c}. Finally, the results are summarized in Sec.~\ref{sec4}.

\section{Methods}
\label{sec:comp}

\subsection{VASP calculations\label{sec2a}}

To model the geometries of Fe atomic chains on the Re(0001) surface, the equilibrium structure of an Fe adatom on Re(0001) has been first calculated by using the VASP method. The obtained structure corresponds to the total energy minimum after geometry optimization. In the calculation the generalized gradient approximation (GGA) within density functional theory (DFT) has been used 
with the exchange--correlation (XC) functional parametrized following the work of Perdew, Burke, and Ernzerhof (PBE) \cite{PBE}. The system has been modeled as a $7\times 7$ surface cell in a slab geometry consisting of four atomic layers of Re (in total $4\times 7\times 7=196$ Re atoms), and an Fe adatom in the hcp hollow position \cite{Kim2018}. The chosen geometry 
ensures that the interactions between Fe atoms in repetitive supercells are negligible due to their large separation of $\sim19.3$\,\AA\: that corresponds to $7\, a_{Re}$, where $a_{Re}=2.761$\,\AA\: is the in-plane lattice constant of Re. In the (0001) direction a 10-\AA-thick vacuum region has been considered to avoid interaction between repetitive slabs. The Brillouin zone was sampled by the Gamma point only due to the large size of the supercell. The Re atoms in the bottom three layers of the slab have been fixed to their hcp bulk positions, and the vertical positions of all Re atoms in the topmost layer and the Fe adatom have been optimized by using a force convergence criterion of 0.01 eV/\AA\: acting on the individual atoms. The Fe adatom is found to have a spin magnetic moment of $2.61\, \mu_{\textrm{B}}$, and it pulls out its three nearest-neighbor (NN) Re atoms slightly from the top Re layer, arriving at a Fe-Re vertical distance of $1.85$\,\AA\: with respect to these nearest neighbors. We also find that the top Re layer relaxes toward the substrate which leads to a vertical Re-Re distance of $2.16$\,\AA\: between the above-mentioned three NN Re atoms of the Fe adatom and the Re atoms in the subsurface layer, which is smaller than the bulk Re interlayer distance of $2.228$\,\AA. These Fe-Re and Re-Re vertical distances were used in the subsequent KKR calculations for the Fe adatom and the atomic chains on Re(0001).

\subsection{KKR calculations\label{sec2b}}

We used the Green's function embedding technique based on the KKR multiple scattering theory \cite{Lazarovits2002} 
to determine the electronic and magnetic properties of the Fe clusters.
The Re(0001) surface has been modeled as an interface region between semi-infinite bulk Re and vacuum consisting of eight atomic layers of Re and four atomic layers of empty spheres (vacuum). The energy integrals were performed using 16 points along a semicircle contour in the upper complex semiplane and a sampling of up to 3282 $\vec{k}$ points in the Brillouin zone was used to calculate the Green's function of the host. The Ceperley--Alder-type of exchange-correlation functionals \cite{Ceperley-Alder-1980} as parametrized by Perdew and Zunger \cite{Perdew-Zunger-1981} and an angular momentum cutoff of $l_{\textrm{max}}=2$ was considered in the KKR calculations, similarly to Ref.~\cite{Laszloffy2017}. A single Fe adatom and chains consisting of five, 10, and 15 Fe atoms were calculated by embedding them in the first vacuum layer with the layer relaxations described in the previous section.

Three different methods have been used to investigate the magnetic properties of the systems. First, the relativistic torque method \cite{Udvardi2003,Ebert2009} was applied to determine parameters of a classical spin model restricted to two-spin interactions. The energies of magnetic configurations within this description were compared by atomistic spin model simulations. The spin model is discussed in Sec.~\ref{sec2c}, while the method for fitting the parameters is given in Appendix~\ref{sec:torque}. Second, the energies of selected spin configurations were compared, such as collinear states with different magnetic orientations or spin spirals with different periods. In the spirit of the magnetic force theorem (MFT) \cite{Liechtenstein1987}, these energy differences between magnetic configurations are calculated with fixed electronic potentials based on the band energy,
which is obtained by using Lloyd's formula \cite{Lloyd1967}. 
Third, the ground state of the Fe chain was also determined completely within the \textit{ab initio} formalism, by updating the spin directions based on the torque acting on them and also performing self-consistent calculations in the obtained spin configurations. This method enables finding local energy minima, ideally the ground state, in the whole configuration space of the spin directions \cite{Balogh2012}. It should be noted that in the second and third methods there is no restriction on the possible types of magnetic interactions apart from those enforced by the symmetry of the system. 

First, we performed self-consistent calculations for an Fe adatom in hcp position on the top of the Re(0001) substrate with the embedded cluster KKR technique. We considered clusters of different sizes and concluded that the spin magnetic moment of Fe, $m_{\rm Fe}=2.46 \, \mu_B$, changes by less than 1~\% when increasing the size of the cluster from 13 lattice sites including three Re atoms and nine empty spheres in the first NN shell to 122 lattice sites including the first three neighbor shells around the Fe adatom. Note that the obtained spin moment of Fe is about 6\% less than the value of $2.61 \, \mu_B$ from the VASP calculations. The magnetic moment of the Fe atom induces a small ($<0.1 \, \mu_B$) magnetic moment in the Re atoms directly below it, while the induced moments of farther Re atoms are negligible. 

We determined the anisotropy energy of the adatom in the spirit of the MFT by calculating the energy difference $\Delta E$ between the cases where the Fe spin is pointing in-plane ($E_\parallel$) and normal to the plane ($E_\perp$). Due to the small value of the induced Re moments the  exchange-correlation field was set to zero at the Re sites while calculating the energy differences, and we obtained that the single Fe adatom has easy-plane anisotropy with $\Delta E = E_\perp - E_\parallel = 0.905\,\text{meV}$. We confirmed that taking into account the exchange-correlation field on the Re sites leads to a change within about 5~\% in the magnetic anisotropy energy; therefore, in all calculations of the Fe chains in terms of the MFT we chose the above approach for simplicity.

We considered close-packed monatomic chains of five, 10,  and 15 Fe atoms along the nearest-neighbor direction on the top of Re(0001). In the following this direction will be denoted by $x$, the in-plane direction perpendicular to $x$ by $y$, and the normal-to-plane direction by $z$. Based on our investigations for the adatom, we considered clusters containing the atomic positions in a NN environment relative to the Fe atoms, including 11, 21, and 31 Re atoms, as well as 25, 45, and 65 empty spheres for the chains of five, 10, and 15 Fe atoms, respectively.

For the five-atom-long Fe chain we first performed self-consistent calculations with ferromagnetic (FM) order and used the torque method to generate a spin model. As will be discussed in Sec.~\ref{sec3a}, the NN isotropic couplings are strongly antiferromagnetic (AFM), implying that an alternating AFM order is considerably lower in energy than the FM state. In order to check the preference for the AFM state, we recalculated the potentials with the AFM order of the spins and,
by using these potentials, we calculated the energy difference between the AFM and FM states within the MFT. Indeed, the AFM state was by 15.7 meV/Fe atom lower in energy than the FM state. 
Based on the above results, for all the Fe chains under consideration we used the self-consistent potentials obtained from alternating AFM configurations to generate the spin-model parameters.

\subsection{Spin model\label{sec2c}}

The adiabatic decoupling of the electronic and spin degrees of freedom and the rigid spin approximation \cite{Antropov1996} make it possible to characterize the energy of a magnetic system by a set of unit vectors $ \left\{ \vrd{e} \right\} \equiv $ $\left\{ \vrd{e}_1, \vrd{e}_2, \dots \vrd{e}_N \right\}$ describing the directions of atomic magnetic moments, where $N$ is the number of magnetic atoms in the system. Since the metallic substrate acts as a particle reservoir for the clusters considered in the calculations, instead of the energy $E$ we will consider the grand potential $\Omega=E-\varepsilon_{\rm F}N_{\rm e}$ at zero temperature, with $\varepsilon_{\rm F}$ and $N_{\rm e}$ being the Fermi energy of the reservoir and the number of electrons in the cluster, respectively. Taking into account one-spin terms and 
 two-spin magnetic interactions, the spin model can be written as 
\begin{equation}
	\Omega\left( \left\{ \vrd{e} \right\} \right) = \Omega_0 + \sum_{i=1}^N \vrd{e}_i \mxrd{K}_i \vrd{e}_i - \frac{1}{2} \sum_{\substack{
   i,j=1 \\
   i \neq j}}^N \vrd{e}_i \mxrd{J}_{ij} \vrd{e}_j,
		\label{eq:heis}
	\end{equation}
	where $\Omega_0$ is a constant, the $ \mxrd{K}_i $ are traceless and diagonal second-order single-ion anisotropy matrices, and the $\mxrd{J}_{ij}$ are tensorial exchange interactions \cite{Udvardi2003}.  The matrices  $\mxrd{J}_{ij}$ can be decomposed into three parts,
\begin{equation}
	\mxrd{J}_{ij} = J_{ij}^I\mxrd{I} + \mxrd{J}_{ij}^S + \mxrd{J}_{ij}^A ,
\end{equation}
	where 
\begin{equation}
	J_{ij}^I =\frac{1}{3} \Tr \left( \mxrd{J}_{ij} \right) \label{eqniso}
\end{equation}
	is the isotropic exchange interaction, 
\begin{equation}
	\mxrd{J}_{ij}^S = \frac{1}{2} \left( \mxrd{J}_{ij} + \mxrd{J}_{ij}^T \right) -J_{ij}^{I} \mxrd{I}
 \end{equation}
is the traceless symmetric part of the matrix, with $T$ denoting the transpose. This is known to contribute to the so-called two-ion magnetic anisotropy of the system. The antisymmetric part of the matrix,
\begin {equation}
	 \mxrd{J}_{ij}^A = \frac{1}{2} \left( \mxrd{J}_{ij}-\mxrd{J}_{ij}^T \right), 
 \end{equation}
is related to the DM interaction \cite{Dzyaloshinsky1958,Moriya1960},
\begin {equation}
 \vrd{e}_i \mxrd{J}^A_{ij} \vrd{e}_j = \vrd{D}_{ij} \left(\vrd{e}_i \times \vrd{e}_j \right)
 \label{eq:dm}
  \end{equation}
  with the DM vector $D_{ij}^\alpha = \frac{1}{2} \varepsilon_{\alpha \beta \gamma} J_{ij}^{\beta \gamma}$, $ \varepsilon_{\alpha \beta \gamma}$ being the Levi--Civita symbol and $\alpha,\beta,\gamma$ denoting Cartesian components.

\begin{figure*}[htb]
\begin{center}
\includegraphics[scale=0.55]{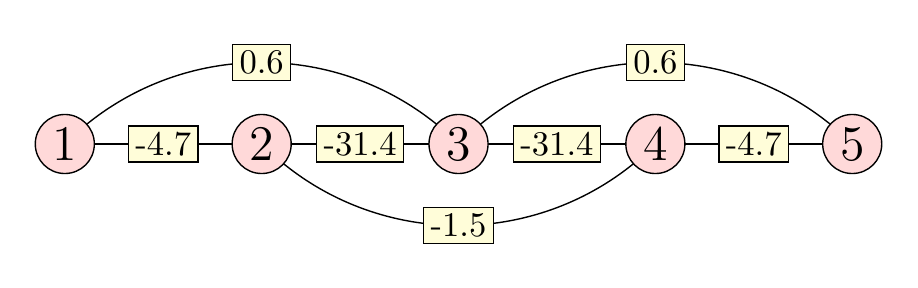}

\includegraphics[scale=0.55]{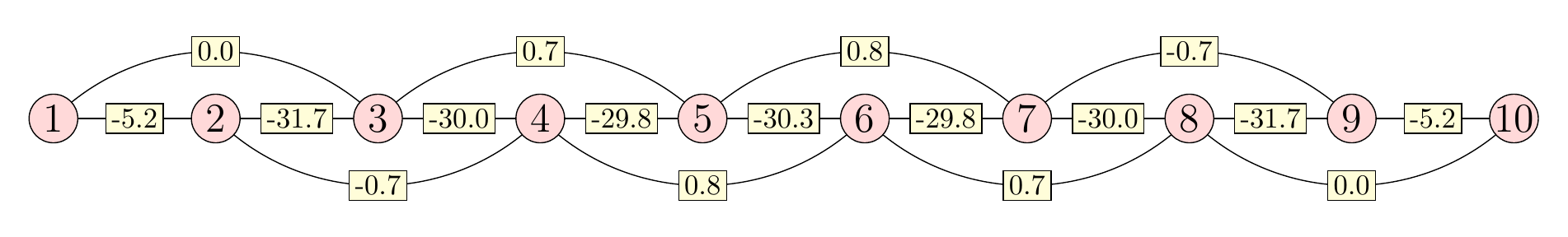}

\includegraphics[scale=0.55]{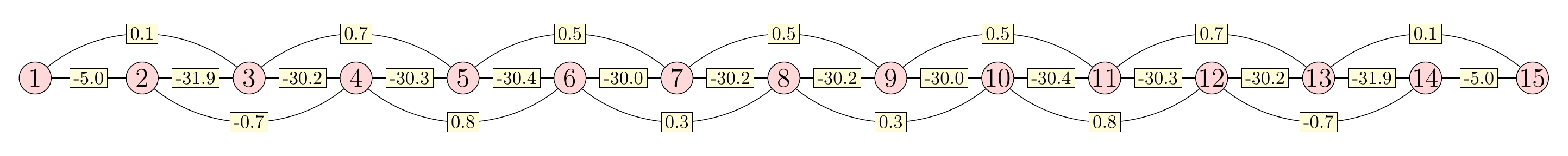}
\caption{Nearest-neighbor (NN) and next-nearest-neighbor (NNN) isotropic interactions $J_{ij}^I$ in the Fe chains of three different lengths.  The interaction strengths between pairs of atoms connected by bonds are presented in the yellow boxes in units of meV. Positive and negative signs correspond to FM and AFM couplings, respectively.}
\label{fig:is}
\end{center}
\end{figure*}   

 Following Ref.~\cite{Laszloffy2017}, site-resolved easy-axis directions and anisotropy energies have been determined for the monatomic chains from the spin model, taking into account both single-ion and two-ion contributions. Since the nearest-neighbor isotropic interactions are found to be antiferromagnetic (see Sec.~\ref{sec3a}), in order to characterize the magnetic anisotropy we consider alternating local moments $\vrd{e}_{i}=(-1)^i \vrd{e}$. 
The grand potential of the system can then be expressed as
\begin{equation}
	\Omega\left( \vrd{e} \right) = \Omega_0^\prime + \sum _{i=1}^N \vrd{e} \mxrd{A}_i \vrd{e} ,
	\label{eq:anitot}
\end{equation}
	with
\begin{equation}
	 \Omega_0^\prime = \Omega_0 - \frac{1}{2} \sum _{i \ne j}  J^{I}_{ij} (-1)^{i+j} ,
	\label{eq:anitot0}
\end{equation}
 and the effective anisotropy matrices
\begin{equation}
 \mxrd{A}_{i} = \mxrd{K}_{i} - \frac{1}{2} \sum\limits_{j=1}^{N} \mxrd{J}_{ij}^{S}(-1)^{i+j}.
 \label{eq:anisite}
\end{equation}

The normalized eigenvectors of the symmetric matrices in Eq.~(\ref{eq:anisite}),  $\vrd{e}^{\,e}_i$, $\vrd{e}^{\,i}_{i}$, and $\vrd{e}^{\,h}_{i}$ correspond in order to the easy, intermediate, and hard directions,  with the respective energy eigenvalues $k^{e}_{i} \leq k^{i}_{i} \leq k^{h}_{i}$. For illustrating the site-specific easy directions together with the magnetic anisotropy energies, we will use the following vector:
\begin{equation}
\vrd{k}_i = \left( k^h_i - k^e_i \right) \vrd{e}^{\,e}_i .
\label{eq:ke}
\end{equation}

The ground state of the spin model was determined 
by zero-temperature Landau--Lifshitz--Gilbert (LLG) spin dynamics simulations where only the damping term was kept. This is described by the time integration step
\begin{equation}
\vrd{e}_{k}^{\,\prime}(t_{n+1})=\vrd{e}_{k}(t_{n})-\lambda \vrd{e}_k(t_{n})\times \left( \vrd{e}_k(t_{n}) \times \vrd{B}^\text{eff}_k(t_{n}) \right),
\end{equation}
where 
\begin{equation}
 \vrd{B}^\text{eff}_{k} = \sum_{j} \mxrd J_{kj} \vrd{e}_{j} - 2\mxrd{K}_{k} \vrd{e}_{k}
\end{equation}
is the effective magnetic field, and a small damping parameter, $\lambda \sim 10^{-4}\,\text{meV}^{-1}$, was chosen. The new spin vectors were normalized after each step to preserve the unit length of the vectors.
The simulations were stopped when the spin components changed less than $10^{-6}$ in $5\cdot 10^5$ subsequent LLG steps.
For each system ten runs with independently chosen random initial configurations were performed which all led to the same final state, providing a strong indication that this is the actual ground state of $\Omega\left( \left\{ \vrd{e} \right\} \right)$ instead of a local minimum.

\section{Results}

\subsection{Spin-model parameters for the Fe chains \label{sec3a}}

In this section we discuss the parameters of the spin model containing two-spin interactions described in Sec.~\ref{sec2c}, calculated in terms of the KKR method and the relativistic torque method detailed in Sec.~\ref{sec2b} and in Appendix~\ref{sec:torque}, respectively. The variation of NN and next-nearest-neighbor (NNN) isotropic interactions $J_{ij}^I$ from Eq.~(\ref{eqniso}) along the chains can be seen in Fig.~\ref{fig:is}.
For all chains, the NN isotropic interactions are the strongest, and their negative sign means AFM coupling. The isotropic interactions become more and more homogeneous at the middle of the chain as the length of the chain is increased. It is also apparent from Fig.~\ref{fig:is}  that the isotropic NN interactions are considerably smaller in magnitude at the edges of the chain than inside the chain for all chain lengths.

\begin{figure}[htb]
\begin{center}
\includegraphics[width=\columnwidth]{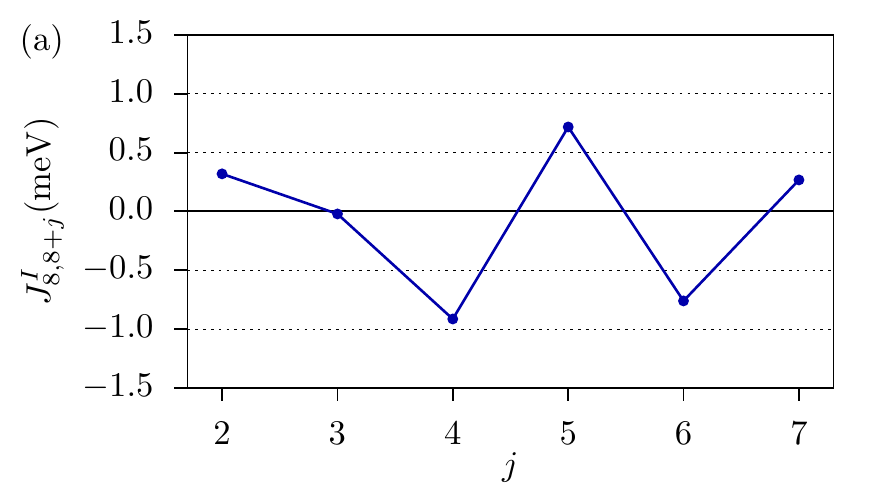}

\includegraphics[width=\columnwidth]{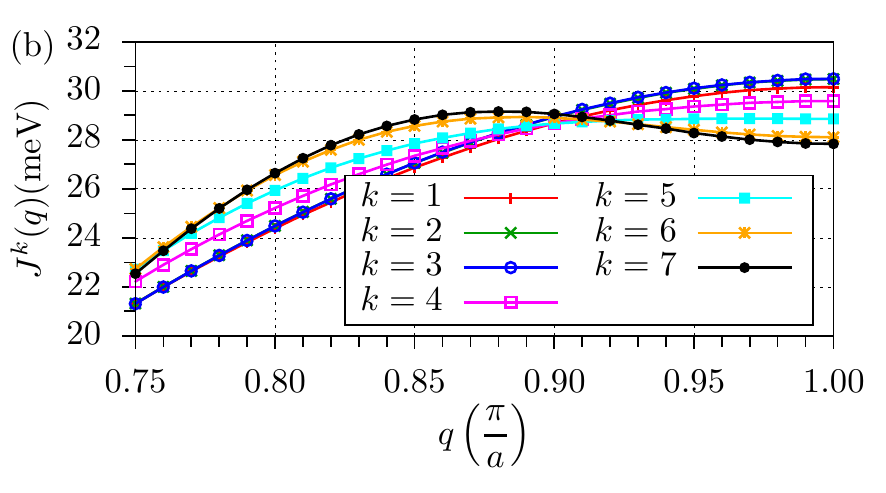}
\end{center}
\vskip -15pt
\caption{(a) Isotropic couplings 
in the 15-atom-long Fe chain between the atom at the middle of the chain indexed by $8$ (see the bottom panel of Fig.~\ref{fig:is}) and the Fe atoms at positions $8+j$ ($j =2,\dots,7$). (b) Fourier transform of the isotropic couplings, Eq.~(\ref{eqnFT}), where $k$ labels the number of neighbors taken into account in the sum.}
\label{fig:fe15_jq}
\end{figure}

The ferromagnetic NNN interactions are more than one order of magnitude smaller than the NN ones as shown in Fig.~\ref{fig:is}. This also holds true for the interactions for farther neighbors as can be seen in Fig.~\ref{fig:fe15_jq}(a), where calculated values for the middle spin $8$ in the 15-atom-long 
chain are displayed. However, if one summarizes the effect of farther interactions, it turns out that they play an important role in determining the ground state. This is demonstrated by introducing the Fourier transform of the isotropic interactions as
\begin{equation}
 J^k(q) = \sum_{j=1}^{k} J_{8,8+j}^I\cos (jaq),~q\in \left[-\frac{\pi}{a},\frac{\pi}{a}\right], \label{eqnFT}
\end{equation}
where $k$ is the number of neighbors taken into account. If one assumes that the interactions in the middle of the chain will no longer be significantly modified as the chain length is increased, then $J^k(q)$ may be used as an approximation for the energy contribution of the isotropic interactions to homogeneous spin spiral states in infinitely long chains, where $q=0$ corresponds to the FM and $q=\frac{\pi}{a}$ to the AFM state. The most favorable state is given by the maximum of $J^k(q)$. It can be seen in Fig.~\ref{fig:fe15_jq}(b) that up to $k=5$ this corresponds to the collinear AFM state. However, considering more shells ($k>5$) in the sum in Eq.~\eqref{eqnFT} a spin spiral state becomes the most favorable, which can be regarded as a long-wavelength modulation of the AFM state. This is a consequence of the frustration of the isotropic interactions, which in the AFM state is indicated by the fact that the isotropic interaction between atoms at the distance of an odd multiple $j$ of the lattice constant becomes FM and at an even $j$ it becomes AFM, although the spins at odd and even $j$ positions should be antiparallel and parallel in the AFM state, respectively. It is shown in Fig.~\ref{fig:fe15_jq}(a) that such kind of frustration occurs for $j \ge 4$, explaining how the spin spiral state is formed as the number of shells is increased.

\begin{figure*}[htb]
\begin{center}
\includegraphics{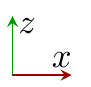}
\includegraphics[scale=0.5]{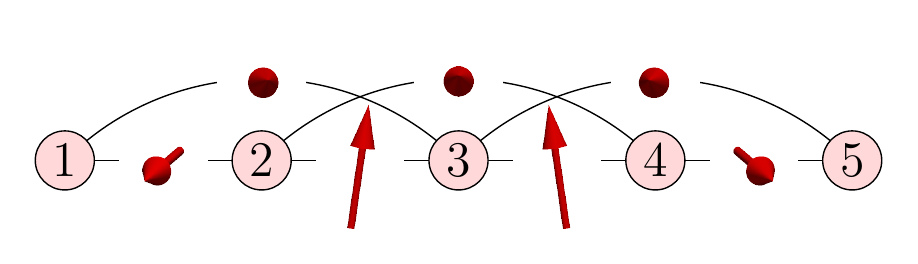}

\includegraphics{marker_side.pdf}
\includegraphics[scale=0.5]{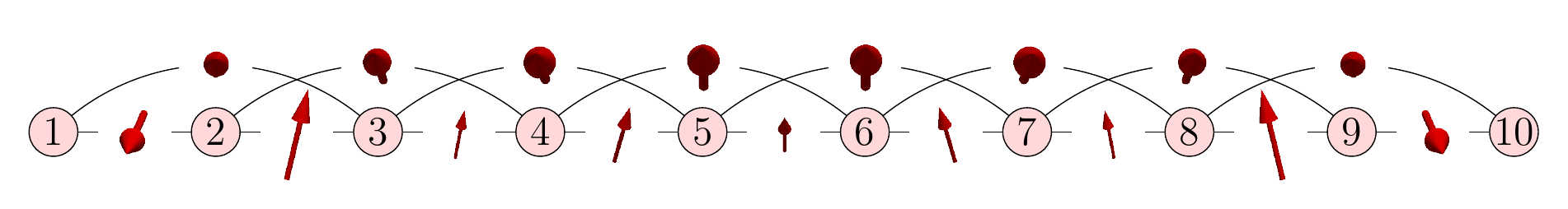}

\includegraphics{marker_side.pdf}
\includegraphics[scale=0.5]{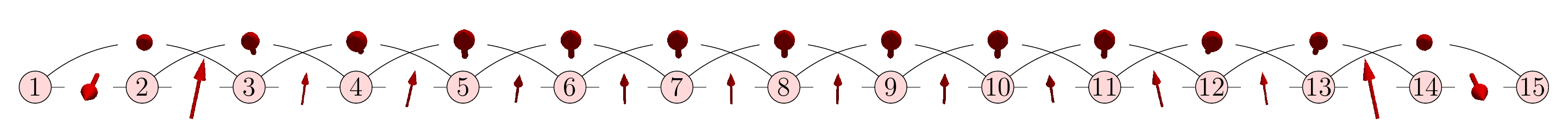}
\caption{Side view of the NN and NNN DM vectors $\vrd{D}_{ij}$ from Eq.~(\ref{eq:dm}) in the Fe chains of three different lengths. With respect to the direction parallel to the chains, the arrows are placed at the centers of the lines between the corresponding pairs of Fe atoms, where at the left side of the line is the atom number $i$ and at the right side is the atom number $j$ in the vector $\vrd{D}_{ij}$. The lengths of the arrows are scaled according to the magnitude of the DM vectors. The numerical values of the components of the DM vectors for the 15-atom-long 
 chain can be found in Table~\ref{table:dm}.}
\label{fig:dm}
\end{center}
\end{figure*}

The NN and NNN DM vectors in the chains, see Eq.~(\ref{eq:dm}), are drawn in Fig.~\ref{fig:dm}. Similarly to the isotropic exchange interactions, the DM vectors at the middle of the chain become stabilized as the chain length is increased.  The DM vectors between the three atoms at the edges of the chains significantly differ from those in the middle of the chains, while these DM vectors are similar between the five-, 10-, and 15-atom-long 
 chains.
The obtained DM interactions satisfy the symmetry rules of Moriya \cite{Moriya1960} with respect to the only crystal symmetry of the system, namely the mirroring at the $yz$ plane intersecting the middle of the chain. Since the DM vector transforms as an axial vector, this symmetry implies
\begin{equation}
 (D_{ij}^{x},D_{ij}^{y},D_{ij}^{z} )= ( D_{\sigma(i),\sigma(j)}^{x},-D_{\sigma(i),\sigma(j)}^{y},-D_{\sigma(i),\sigma(j)}^{z} ) ,
 \label{eqn:mirroring}
 \end{equation}
where $\sigma(i)=N+1-i$ is the mirror image of site $i$ in the chain of length $N$. Note that in Fig.~\ref{fig:dm} the vectors are displayed for $i<j$ and $\vrd{D}_{ij}=-\vrd{D}_{ji}$ by definition.

In the case of the 15-atom-long chain, the numerical values for the components of the NN and NNN DM vectors are given in Table \ref{table:dm}. It can be seen that the NNN DM vectors are the largest in magnitude and they are almost parallel to the $-y$ direction.
Although the $y$ components of the NN DM vectors have the same sign, these are actually competing with the NNN vectors due to the short-range AFM order (see Sec.~\ref{sec3b}), similarly how the alternating isotropic interactions in Fig.~\ref{fig:fe15_jq}(a) are competing with the NN interaction. 
The $z$ components of the NN and NNN DM vectors are mostly positive; they only change sign for the NN atoms at the edge of chain ($D^z_{12}<0$). 
In general, $D^z_{ij}$ is larger for the NNs than for the NNNs, but at the middle of the chain they become roughly similar in size. 
The $x$ components of the DM vectors are very small and they should disappear in the limit of infinitely long chains due to the mirror symmetry with respect to the $yz$ plane. For the 15-atom-long chain, the mirror symmetry implies $D^x_{79}=0$.

\begin{table}[htb]
\begin{ruledtabular}
 \caption{ \label{table:dm} Components of the NN and NNN DM vectors, $\vrd{D}_{i,i+1}$ and $\vrd{D}_{i,i+2}$, respectively, for the 15-atom-long 
 chain, given in units of meV. The components of the DM vectors
 for $i>7$ can be obtained by the symmetry relations Eq.~\eqref{eqn:mirroring} and are illustrated in Fig.~\ref{fig:dm}.
 }
 \begin{tabular}{crrrrrrr}
  &$\vrd{D}_{12}$&$\vrd{D}_{23}$&$\vrd{D}_{34}$&$\vrd{D}_{45}$&$\vrd{D}_{56}$&$\vrd{D}_{67}$&$\vrd{D}_{78}$\\
 \hline
 $x$&--0.60&  0.64&  0.23&  0.42&  0.18&--0.03&--0.03\\
 $y$&--3.95&--0.71&--0.98&--1.24&--1.76&--1.62&--1.43\\
 $z$&--1.24&  2.96&  1.61&  1.80&  1.34&  1.48&  1.49\\ \hline \\ [-5pt]
  &$\vrd{D}_{13}$&$\vrd{D}_{24}$&$\vrd{D}_{35}$&$\vrd{D}_{46}$&$\vrd{D}_{57}$&$\vrd{D}_{68}$&$\vrd{D}_{79}$\\
 \hline
 $x$&  0.05&--0.25&--0.30&--0.04&--0.02&--0.07&  0.00\\
 $y$&--4.06&--4.47&--5.12&--5.13&--4.92&--4.96&--5.03\\
 $z$&  0.44&  0.84&  0.73&  1.16&  1.16&  1.13&  1.12\\
 \end{tabular}
\end{ruledtabular}
\end{table}

\begin{figure*}[htb]
\begin{center}
\includegraphics[scale=1]{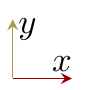}\includegraphics[scale=0.6]{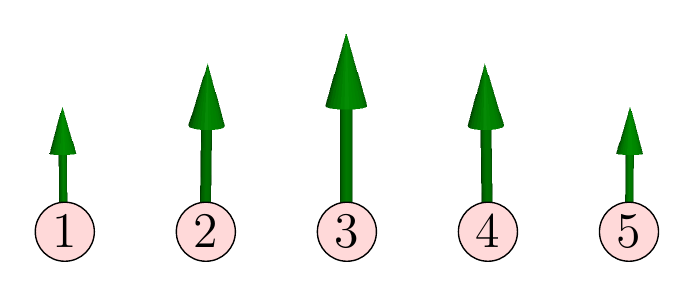}

\includegraphics[scale=1]{marker_top.pdf}\includegraphics[scale=0.6]{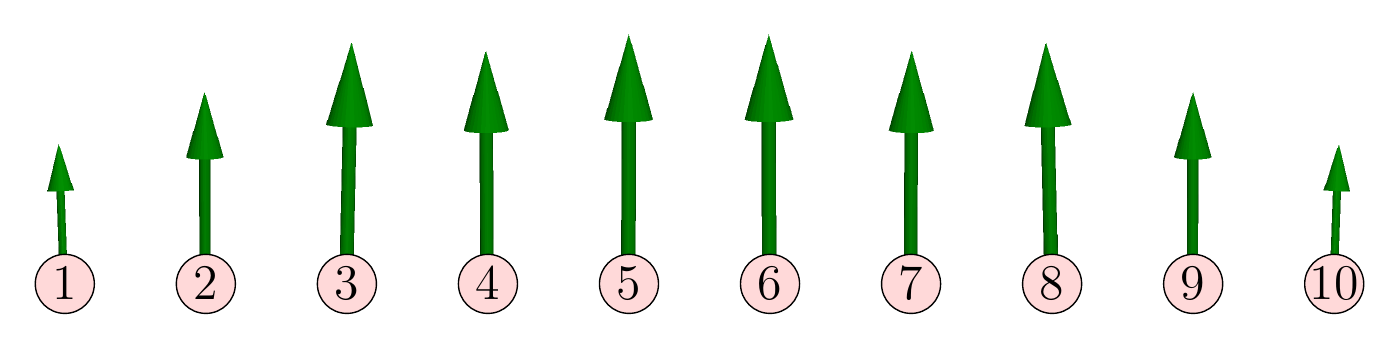}

\includegraphics[scale=1]{marker_top.pdf}\includegraphics[scale=0.6]{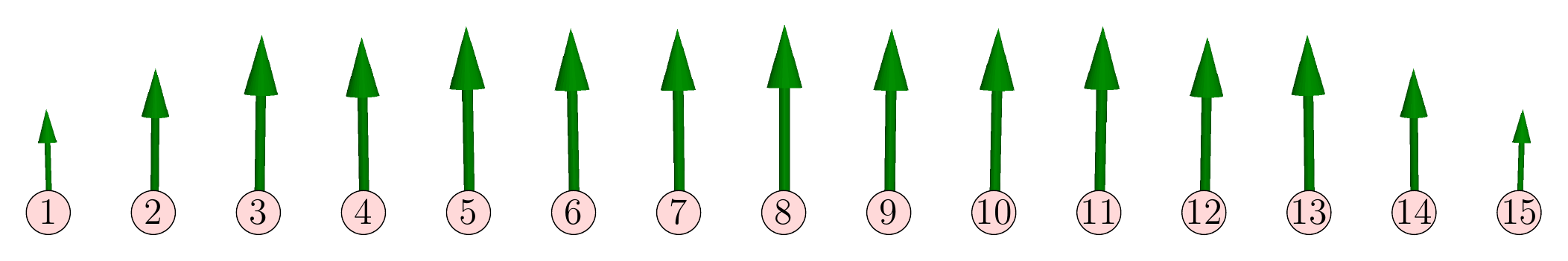}
\caption{Site-resolved anisotropy vectors  according to Eq.~\eqref{eq:ke}. The length of the arrows scales with the magnitude of the anisotropy vectors. The largest magnitude of $\vrd{k}_i$  is 7.79, 9.03, and 8.92~meV for the five-, 10-, and 15-atom-long 
 chains, respectively. 
}
\label{fig:ani}
\end{center}
\end{figure*}

The site-resolved anisotropy vectors defined in Eq.~\eqref{eq:ke} are visualized in Fig.~\ref{fig:ani}, where the arrows point along the easy directions and their magnitude is proportional to the energy difference between hard and easy axes at the given site. At all sites the easy axis is almost parallel to the $y$ direction, while the hard axis is roughly along the $z$ direction. With increasing chain length the magnitude of the
$\vrd{k}_i$ vectors gets quite homogeneous with the maxima of 7.79, 9.03, and 8.92~meV at the middle of the five-, 10-, and 15-atom-long 
 chains, respectively. However, since the site-resolved magnetic anisotropy energy drops at the edge of the chains, the average length of the
anisotropy vectors is 6.25, 7.60, and 7.91 meV for the three chains in order. These values are close to the average anisotropy energies, 
$(\Omega(\hat{z})-\Omega(\hat{y}))/N$ calculated from Eq.~\eqref{eq:anitot}, 5.79, 7.37, and 7.76 meV/Fe, respectively, since as noted the local easy and hard axes are very close to the $y$ and $z$ axes for all the Fe atoms in the chains.

\subsection{Ground state\label{sec3b}}

\begin{figure*}
\begin{center}
\subfigure[]{
\includegraphics[scale=1]{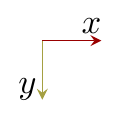}
\includegraphics[width=0.3\textwidth]{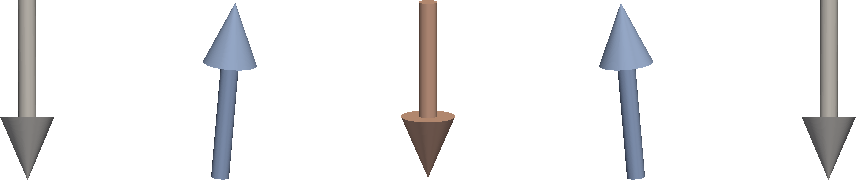}\label{fig:gs_5}}

\subfigure[]{
\includegraphics[scale=1]{marker_top_plusz_v2.pdf}
\includegraphics[width=0.55\textwidth]{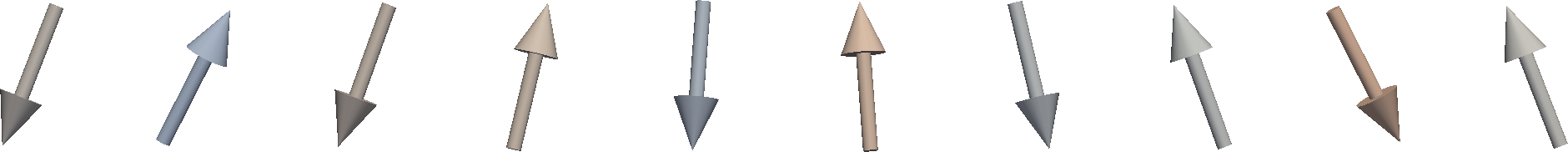}\label{fig:gs_10}}

\subfigure[]{
\includegraphics[scale=1]{marker_top_plusz_v2.pdf}
\includegraphics[width=0.7\textwidth]{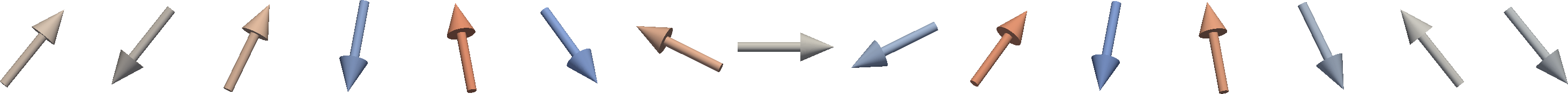}\label{fig:gs_15}}

\subfigure[]{
\includegraphics[scale=1]{marker_top_plusz_v2.pdf}
\includegraphics[width=0.7\textwidth]{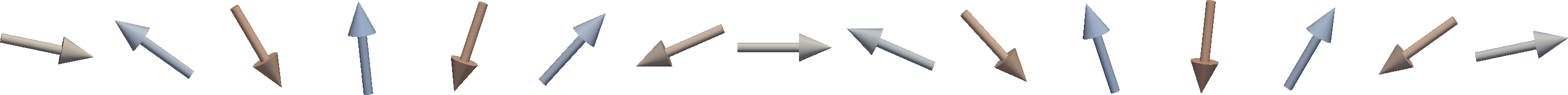}\label{fig:gs_spiralmap}}

\subfigure[]{
\includegraphics[scale=1]{marker_top_plusz_v2.pdf}
\includegraphics[width=0.7\textwidth]{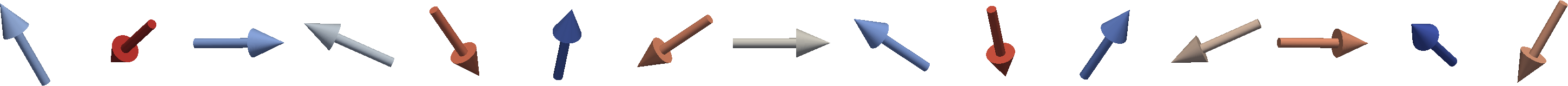}\label{fig:gs_ab}}

\includegraphics{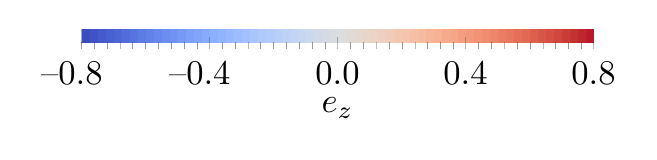}
\caption{Top view of the ground state spin configurations of the chains obtained for different chain lengths and by using different calculation methods: (a) 5 Fe, spin model, (b) 10 Fe, spin model, (c) 15 Fe, spin model, (d) 15 Fe, homogeneous spin spirals, and (e) 15 Fe, \textit{ab initio} spin dynamics. The $z$ component of the normalized spin vectors is visualized by color coding according to the color bar below the figures.} 
\label{fig:gs}
\end{center}
\end{figure*}
 
Here the ground states of the magnetic clusters will be discussed, focusing on the chirality of spin rotation inside the structures. We will use the convention that the rotation of the spins in the $xz$ plane is locally right handed 
 at site $i$ if the projection of $\vec{e}_{i+1}$ on this plane may be obtained from the projection of $\vec{e}_{i}$ via a right-handed rotation by an angle smaller than $180^{\circ}$ around  the positive $y$ axis, meaning that the angle between the projections is smaller than $180^{\circ}$ when rotating from the positive $z$ towards the positive $x$ direction. The opposite chirality is called 
 left handed, 
in agreement with previous definitions of the chirality for spin spirals close to the ferromagnetic state in Ref.~\cite{Heide}. Analogously, we call the rotation in the $xy$ plane locally 
 right handed 
 if the projections of the spins rotate from the positive $x$ towards the positive $y$ direction.

First we determined the ground states of the chains from the spin model following the method described in Sec.~\ref{sec2c}. The obtained configurations can be seen in Figs.~\ref{fig:gs_5}-\ref{fig:gs_15}. The five-atom-long 
 chain prefers AFM ordering due to the strong AFM coupling between the NN spins. 
The $z$ component of the DM vectors is responsible for a slight noncollinearity in the AFM state.
The negative $D_{12}^z$ value in Fig.~\ref{fig:dm} causes a slight left-handed  
rotation between spins 1 and 2, while the positive $D_{23}^z$ causes a right-handed   
rotation between spins 2 and 3.
 The ground state is tilted away from the $xy$ plane through a rotation around the $x$ axis by about 12.8$^\circ$, so the spins with positive $y$ component now have positive $z$ component, too. The tilting is caused by the competition of the easy $y$ axis anisotropy (see Fig.~\ref{fig:ani})
 and the large $y$ components of the DM vectors (similar to those in Table~\ref{table:dm}) preferring a rotation of the spins in the $xz$ plane. 

The ground state of the 10-atom-long 
 chain in Fig.~\ref{fig:gs_10} is again almost collinear AFM due to the strong NN AFM coupling between the spins. 
The $z$ components of the DM vectors shown  in Fig.~\ref{fig:dm}
 cause a rotation of the spins around the $z$ axis along the whole chain, but no full period of a spin spiral state can be observed.
 The ground state is now tilted from the $xy$ plane through a rotation around the $x$ axis by $-6.2^{\circ}$. 

In both the five- and 10-atom-long chains, the strong anisotropy confines the systems close to the $xy$ plane. The chirality is right handed over the middle three atoms in the five-atom-long chain and over the middle eight atoms in the 10-atom-long chain. Note that only looking at every second spin in the chain, this visually corresponds to a left-handed 
rotation, since the angle between the neighboring spins is close to $180^{\circ}$ because of the strong AFM NN couplings. This chirality is determined by the $z$ components of the DM vectors shown in Fig.~\ref{fig:dm}. The direction of the tilting is defined by the chirality in the $xy$ plane  on the one hand and the $y$ components of the DM vectors on the other hand, the latter influencing the rotational sense in the $xz$ plane.
In both systems the NN and NNN DM vectors are characterized by large negative $y$ components, both of which would prefer a left-handed 
rotation if the angle between the NN spins would be small. However, since this angle is larger than $90^\circ$ in the present case, the apparent left-handed rotation between the NNN spins actually corresponds to a right-handed rotation between the NN spins, meaning that the NN and NNN DM vectors are competing in this AFM spin structure. In the five-atom-long chain the rotational plane is tilted around the $x$ axis by a positive angle, leading to a left-handed 
chirality in the $xz$ plane enforced by the NN DM vectors. For the longer chain length of $10$ atoms with the same chirality in the $xy$ plane the influence of the NNN DM vectors becomes stronger, leading to a tilting around the $x$ axis by a negative angle and a right-handed 
chirality in the $xz$ plane.

The ground state of the 15-atom-long 
 chain shown in Fig.~\ref{fig:gs_15} can much better be characterized as a spin spiral state. As described in the context of Fig.~\ref{fig:fe15_jq}, 
the frustrated isotropic interactions induce a spin spiral with a wave number of $q_{\text{max}} = 0.88 \frac{\pi}{a}$ (see the $k=7$ curve), 
which corresponds to a wavelength of about $17 \, a$, slightly larger than the length of the chain. In Fig.~\ref{fig:gs_15} the spins from positions 4 to 11 visually form a full period,
which can be understood as a $14 \, a$ wavelength modulation of the AFM state.
This shorter modulation period might easily be caused by the $z$ components of the NN DM vectors, which are not included in Fig.~\ref{fig:fe15_jq}. The large negative $y$ components of the NNN DM vectors play an important role in tilting the rotational plane out from the $xy$ plane by $-26.1^{\circ}$ around the $x$ axis, for a simple explanation see Appendix~\ref{sec:tilted}. A similar tilted spin spiral state was already attributed to the interplay of the DM interaction and easy-plane magnetic anisotropy in Ref.~\cite{Simon2018}. Similarly to the 10-atom-long chain, the DM vectors imply right-handed 
rotation both in the $xy$ plane and in the $xz$ plane over the whole chain.  
This ground state cannot satisfactorily be reconciled with the experimental observation of a four-atomic period in a chain of 40 Fe atoms reported in Ref.~\cite{Kim2018}, which would correspond to a spin spiral state where neighboring spins are perpendicular to each other.
 
\begin{figure}[h!]
\begin{center}
\includegraphics[width=\columnwidth]{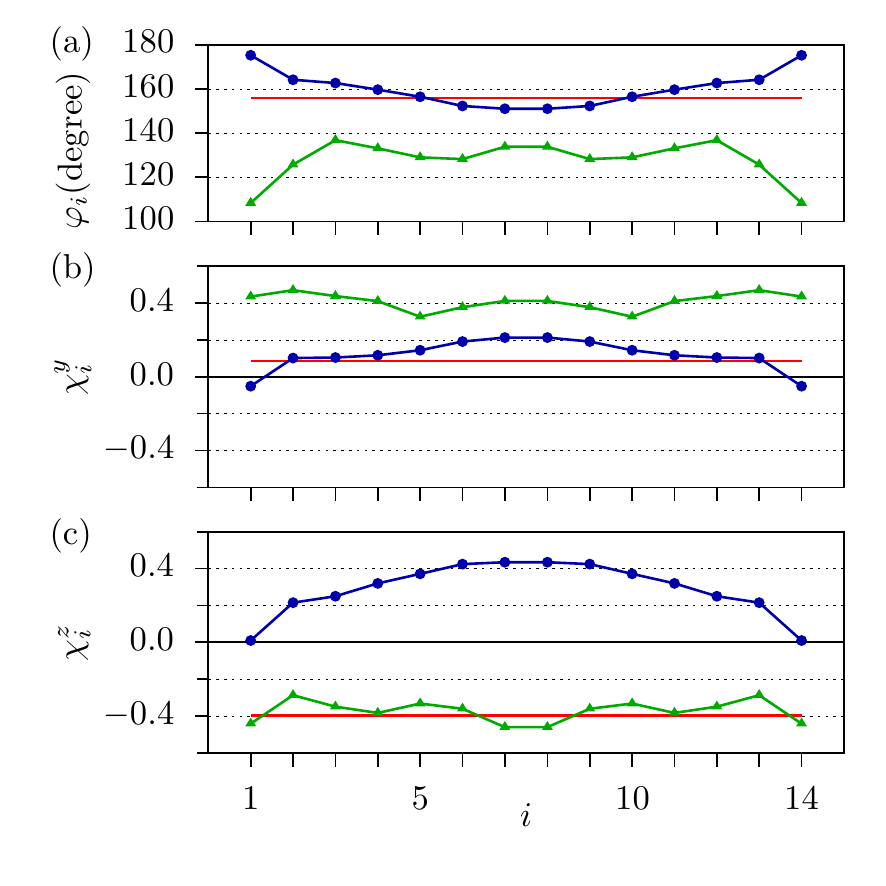}
\end{center}
\vskip -15pt
\caption{(a) The angles between NN spins $\varphi_{i}$, as well as (b) the $y$ and (c) the $z$ components of the chirality vectors $\vrd{\chi}_{i}$ in the 15-atom-long Fe chain for the three different calculation methods; blue circles: spin model, green triangles: \textit{ab initio} spin dynamics, red horizontal lines: homogeneous spin spirals.
 }
\label{fig:fe15_tilt}
\end{figure}

The local rotational sense of the spins in a spin spiral discussed above can be quantitatively described by the site-dependent chirality vector defined as 
\begin{equation}
  \vrd{\chi}_i = \vrd{e}_i \times \vrd{e}_{i+1} \, .
\end{equation}
If only NN DM vectors were included in the model, $\vrd{\chi}_i$ would be parallel to the direction of $\vrd{D}_{i,i+1}$. The chirality is right-handed  
in the $xy$ plane and in the $xz$ plane for $\chi_{i}^{z}>0$ and $\chi_{i}^{y}>0$, respectively.
In Fig.~\ref{fig:fe15_tilt} the magnitude of the local chirality vectors $\left| \vrd{\chi}_{i} \right|=\sin\varphi_{i}$, where $\varphi_{i} \in \left[ 0^{\circ}, 180^{\circ} \right]$ is the angle between the NN spins, as well as the $y$ and $z$ components of the chirality vectors are displayed for the 15-atom-long chain. The angle between the spins at the edges of the chain is almost $180^{\circ}$ and also the sign of $\chi_{i}^{y}$ is switched compared to the middle of the chain, which can be explained by the edge effects in the interaction parameters discussed in Sec.~\ref{sec3a}. 
In the middle of the chain the local environment of the sites is similar, leading only to slight variations in $\varphi_i$.  
The $y$ component of the chirality vectors takes a value between 0.1 and 0.2 for the spins 2 to 13, indicating that the spins tilt away from the $xy$ plane as argued in Appendix~\ref{sec:tilted}.

For comparison with the ground state obtained from the spin model, we calculated the energy of the 15-atom-long Fe chain in the homogeneous spin spiral configuration,
\begin{equation}
\begin{aligned}
 \vrd{e}_i = &\left( \cos \left[ i\left( \pi + \delta \right) + \varphi_0 \right],\cos\alpha \sin \left[ i\left( \pi +\delta \right) +\varphi_0 \right] , \right. \\ 
 &\left. \sin\alpha \sin\left[ i\left( \pi +\delta \right) + \varphi_0 \right] \right) \, ,
\end{aligned}
\label{eq:spiralstate}
\end{equation}
in the spirit of the MFT by keeping the  AFM potentials fixed.
Here $\delta$ is the modulation angle of the AFM state, $\alpha$ is the tilting angle of the spiral from the $xy$ plane and $\varphi_0$ is a phase factor. The NN spin angle is $\varphi = \pi + \delta $, indicating right-handed 
rotation in the $xz$ and left-handed 
rotation in the $xy$ planes for $\delta>0^{\circ}$ and $0^{\circ}\le \alpha \le 90^{\circ}$, respectively. Both rotational senses switch under a sign change of $\delta$, and the rotation in the $xy$ plane proceeds in the opposite direction for $90^\circ\le\alpha \le 180^\circ$. We did not consider an additional angle variable which would differentiate between
left- and right-handed 
rotational senses in the $yz$ plane, since in an infinitely long chain these two chiralities are equivalent due to the mirror symmetry with respect to the $yz$ plane. The phase factor $\varphi_0$ was determined by minimizing the anisotropy energy assuming a  homogeneous magnetic anisotropy with $y$ easy axis, i.e., maximizing $\sum_{i=1}^N \sin ^2 \left[ i\left( \pi +\delta \right) +\varphi_0 \right] $ for every $\delta$.

\begin{figure}[htb]
\begin{center}
\includegraphics[width=\columnwidth]{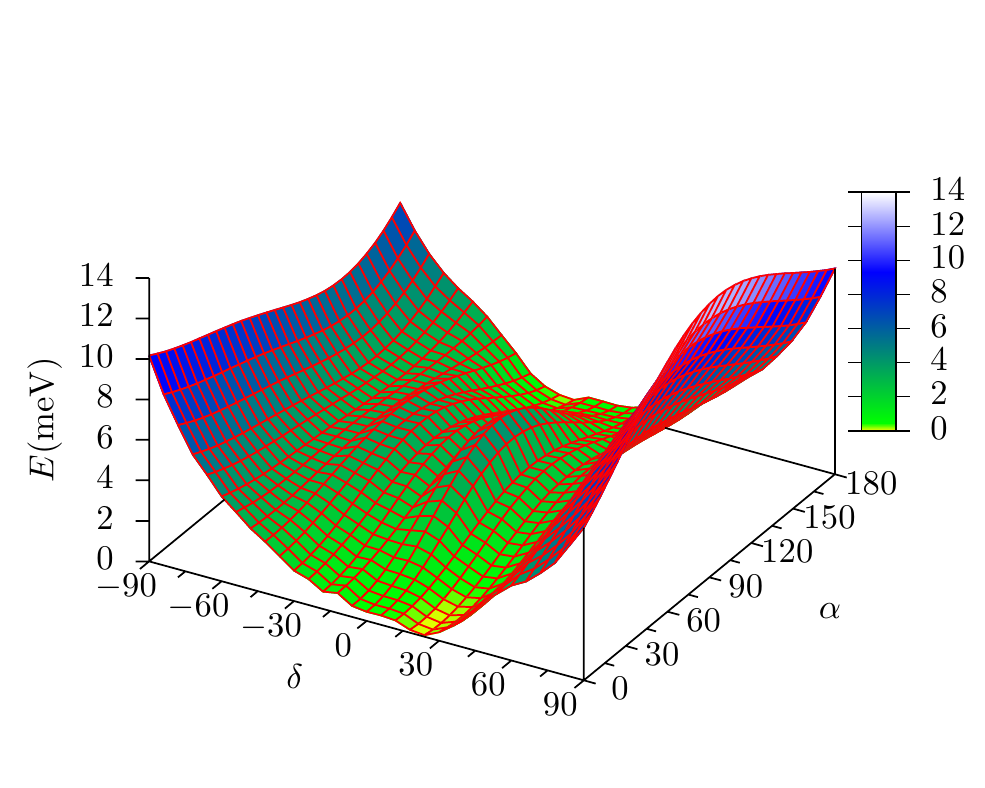}
\end{center}
\caption{Energy per atom of different spin spiral configurations according to Eq.~(\ref{eq:spiralstate}), calculated within the MFT for the 15-atom-long chain. $\delta$ denotes the NN angle of the spiral with respect to the collinear AFM state and $\alpha$ is the tilting angle from the $xy$ plane. 
The minimum, set as the zero level of the energy, is found at $\delta=24^\circ$, $\alpha=12^\circ$. }
\label{fig:spiralmap}
\end{figure}

The energy of the spin spirals normalized to one Fe atom is shown in Fig.~\ref{fig:spiralmap} as a function of $\delta$ and $\alpha$, where the zero level corresponds to the state with the lowest energy. The obtained minimum is at $\delta=24^\circ$ and $\alpha=12^\circ$, with the corresponding spin configuration shown in Fig.~\ref{fig:gs_spiralmap}. While the period of this spin spiral, $\lambda \simeq 15 \, a$, shows good agreement with that obtained from the simulations based on the spin model shown in Fig.~\ref{fig:gs_15}, the actual values of $\delta$ and $\alpha$
indicate a right-handed 
rotation in the $xz$ and a left-handed  
rotation in the $xy$ planes (see above), the latter being the opposite of the spin model simulation results. This can clearly be seen in Fig.~\ref{fig:fe15_tilt}(c), where $\chi^z_{i}$ for the ground state of the spin model and for the spin spiral with lowest energy are opposite in sign.

In the case of the 15-atom-long chain we also performed a fully \textit{ab initio} spin dynamics energy minimization as described in Sec.~\ref{sec2b}.
The energetically most favorable configuration found by this method is shown in Fig.~\ref{fig:gs_ab}, which also resembles a flat spin spiral state apart from small deviations from the coplanar spin arrangement. This method predicts a right-handed rotation of the spins in the $xz$ and a left-handed rotation in the $xy$ planes, in agreement with the MFT calculations performed for the homogeneous spin spiral states. This approach results in a wavelength of $\lambda \simeq 10 \, a$ ($\delta \simeq 36^{\circ}$) of the spin spiral modulation which is significantly shorter than those obtained from the previous two methods. On top of the AFM state, this visually leads to a $5 \, a$ period of the spin structure, see spins 3, 8, and 13 in Fig.~\ref{fig:gs_ab}, which fits the experimental observation the most.

\subsection{Four-spin chiral interactions\label{sec3c}}

\begin{figure}
\begin{center}
\includegraphics[scale=1]{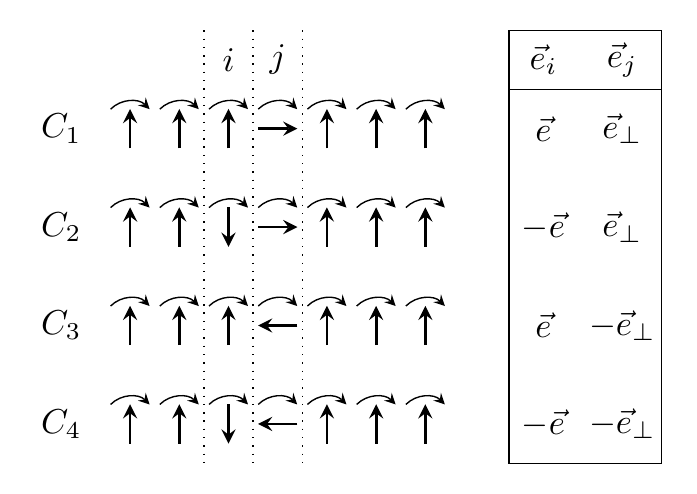}
\end{center}
\caption{Illustration of the states for determining chiral interactions from MFT calculations. In each configuration $C_{1}$-$C_{4}$, a global rotation of the spins around the $y$ direction is performed, described by the vectors $\vrd{e}=\left( \sin\theta , 0, \cos\theta\right)$ and $\vrd{e}_\bot = \left( \cos \theta ,0,-\sin \theta \right)$.}
\label{fig:initstates}
\end{figure}

It was found in Sec.~\ref{sec3b} that the spin model Eq.~\eqref{eq:heis} and the MFT calculation of homogeneous spin spirals yield opposite rotational senses of the spin components in the $xy$ plane for the 15-atom-long chain. This most likely indicates that multispin interactions play an important role in the present system, since these were not taken into account in the spin model, but are implicitly included in the MFT calculations. In order to estimate the magnitude of these interactions, it is worthwhile to calculate the chiral interactions from energy differences based on the MFT directly and compare them to the values of the spin model shown in Table~\ref{table:dm}. This method is illustrated for the $y$ component in Fig.~\ref{fig:initstates}, where four configurations in the $xz$ plane are shown.
In the configuration denoted by $C_{1}$, during a global rotation of the spins around the $y$ axis by the angle $\theta$ most spins are rotated according to $\vrd{e}=\left( \sin\theta , 0, \cos\theta\right)$, while the perpendicular spin at site $j$ will follow $\vrd{e}_\bot = \left( \cos \theta ,0,-\sin \theta \right)$. The configurations $C_{2},C_{3},$ and $C_{4}$ are obtained by switching the signs of spins $i$ or $j$ as illustrated on the right side of Fig.~\ref{fig:initstates}.

Assuming the spin model containing only two-spin interactions, from the grand potentials associated to the configurations $C_{1}$-$C_{4}$ one obtains
\begin{equation}
\begin{aligned}
 &\Omega\left(C_1\right)-\Omega\left(C_2\right)-\Omega\left(C_3\right)+\Omega\left(C_4\right) = -2\left( J_{ij}^{zx} - J_{ij}^{xz} \right)\\
 &+ 2 \left( J_{ij}^{xz} + J_{ij}^{zx} \right) \cos 2\theta + 2 \left( J_{ij}^{xx} + J_{ij}^{zz} \right) \sin 2\theta
 \label{eq:bandfitb}
\end{aligned}
\end{equation}
as a function of the rotation angle $\theta$. Note that due to the choice of configurations and the switching of the spin directions only the interaction between sites $i$ and $j$ remains in Eq.~(\ref{eq:bandfitb}). 
Averaging Eq.~\eqref{eq:bandfitb} over the angle $\theta$ we define the $y$ component of the chiral interaction vector as
\begin{equation}
\mathcal{D}^{y}_{\text{r},ij} \equiv -\frac{1}{4}\left<\Omega\left(C_1\right)-\Omega\left(C_2\right)-\Omega\left(C_3\right)+\Omega\left(C_4\right)\right>,\label{eq:enerchiral}
\end{equation}
which, by comparing with Eq.~\eqref{eq:dm},  corresponds to the $y$ component of the DM vector $\mathcal{D}^{y}_{\text{r},ij}=D_{ij}^{y}=\frac{1}{2} (J_{ij}^{zx} - J_{ij}^{xz})$, if only two-spin interactions are considered in the spin model. Here the index $\text{r}$ denotes that the chiral interaction vector was obtained from the rotational scheme depicted in Fig.~\ref{fig:initstates}. Analogously, the quantities $\mathcal{D}^{x}_{\text{r},ij}$ and $\mathcal{D}^{z}_{\text{r},ij}$ corresponding to the other two components of the DM vector can be calculated by performing the rotation in the $yz$ and $xy$ planes, respectively.

The calculated values for $\mathcal{D}^{\alpha}_{\text{r},ij}$ ($\alpha=x,y,z$) are collected in Table~\ref{tab:4spin} for NN and NNN spins in the middle of the 15-atom-long chain. In addition, the $\mathcal{D}_{\text{t},ij}^{\alpha}$ values obtained from the torque method, defined in Eq.~\eqref{eqnS12} in Appendix~\ref{sec:torque}, are listed in the table, which within the spin model Eq.~\eqref{eq:heis} should also coincide with $D_{ij}^{\alpha}$ (cf. Table~\ref{table:dm}). Due to the mirror symmetry with respect to the $yz$ plane going through atom $8$, the symmetry rules for the DM vectors, Eq.~\eqref{eqn:mirroring},  imply $\left(\mathcal{D}_{\text{t},78}^{x},\mathcal{D}_{\text{t},78}^{y},\mathcal{D}_{\text{t},78}^{z}\right)=\left(-\mathcal{D}_{\text{t},89}^{x},\mathcal{D}_{\text{t},89}^{y},\mathcal{D}_{\text{t},89}^{z}\right)$ and $\mathcal{D}_{\text{t},79}^{x}=0$. Remarkably, these symmetry relations do not apply to the corresponding chiral interaction vectors obtained from the rotational method; in particular, $\mathcal{D}_{\text{r},79}^{x}$ does not vanish, but it has a comparable value to the other components. This means that the spin model parametrization of the band energy surface of the present system is not compatible by taking into account two-spin DM interactions only.

\begin{table}[htb]
\caption{Chiral interaction vectors between pairs of spins obtained from the rotational scheme based on MFT calculations, $\mathcal{D}_{\text{r},ij}^{\alpha}$, Eq.~\eqref{eq:enerchiral}, and from the torque method, $\mathcal{D}_{\text{t},ij}^{\alpha}$, defined in Eq.~\eqref{eqnS12}. Values are given in meV for NN and NNN interactions in the middle of the 15-atom-long chain.}
\begin{ruledtabular}
\begin{tabular}{cccrr}
$i$&$j$ & $\alpha$ & $\mathcal{D}_{\text{r},ij}^{\alpha}$ & $\mathcal{D}_{\text{t},ij}^{\alpha}$ \\\hline
7&8 & $x$ &--0.243&--0.033\\
7&8 & $y$ &--3.107&--1.428\\
7&8 & $z$ &--2.306&  1.491\\
7&9 & $x$ &--0.376&  0.000\\
7&9 & $y$ &--1.083&--5.030\\
7&9 & $z$ &--1.319&  1.117\\
8&9 & $x$ &--0.167&  0.033\\
8&9 & $y$ &--3.154&--1.428\\
8&9 & $z$ &--2.313&  1.491\\
\end{tabular}

\label{tab:4spin}
\end{ruledtabular}
\end{table}

In order to explain the differences between $\mathcal{D}_{\text{r},ij}^{\alpha}$ and $\mathcal{D}_{\text{t},ij}^{\alpha}$ in Table \ref{tab:4spin}, it is necessary to include multispin chiral interactions in the model description. Consider a grand potential of the form
\begin{equation}
\begin{aligned}
 \Omega = &\sum_{i} \vrd{e}_i \mxrd{K}_{i} \vrd{e}_i - \frac{1}{2} \sum_{i,j} \vrd{e}_i \mxrd{J}_{ij} \vrd{e}_j \\
 & - \frac{1}{2}\sum_{i,j}\vrd{D}_{ijij}\left( \vrd{e}_{i} \vrd{e}_{j} \right)\left( \vrd{e}_{i} \times \vrd{e}_{j} \right) \\
 & - \sum_{i,j,k} \vrd{D}_{ijjk} \left( \vrd{e}_{i} \vrd{e}_{j} \right) \left( \vrd{e}_{j} \times \vrd{e}_{k} \right) \\
 & - \frac{1}{4} \sum_{i,j,k,l} \vrd{D}_{ijkl} \left( \vrd{e}_{i} \vrd{e}_{j} \right) \left( \vrd{e}_{k} \times \vrd{e}_{l} \right),
 \label{eq:Heis4}
\end{aligned}
\end{equation} 
where the last three terms represent two-, three-, and four-site four-spin chiral interactions combining isotropic (scalar product) and DM (cross product) contributions. In the sums in Eq.~(\ref{eq:Heis4}), the $i$, $j$, $k$, and $l$ indices run over all lattice sites, and in each sum the different indices label different atoms. Equation~(\ref{eq:Heis4}) may be rewritten in an alternative way where the summations are performed over all pairs, three-site and four-site clusters; this is presented in Appendix~\ref{sec:fourspin}.

The two-site four-spin chiral interactions were recently investigated in Ref.~\cite{Brinker}. Following from the definition Eq.~(\ref{eq:Heis4}), they are antisymmetric in their indices,
\begin{align}
\vrd{D}_{ijij} = -\vrd{D}_{jiji},\label{twosite}
\end{align}
similarly to the two-spin Dzyaloshinsky--Moriya interaction in Eq.~(\ref{eq:dm}). For the three-site interactions we will consistently use the notation $\vrd{D}_{ijjk}$ where the second and third site indices coincide, in which case there is no intrinsic symmetry relation connecting the coefficients in the sum in Eq.~(\ref{eq:Heis4}). By definition, the four-site interactions $\vrd{D}_{ijkl}$ satisfy the symmetry relations
\begin{align}
 \vrd{D}_{ijkl} &= \vrd{D}_{jikl}, \\
 \vrd{D}_{ijkl} &= -\vrd{D}_{ijlk};\label{foursite}
\end{align}
therefore, a prefactor of $1/4$ is introduced in the last term of Eq.~(\ref{eq:Heis4}). 
 Moreover, the mirror symmetry on the $yz$ plane in the center of the chain implies
\begin{equation}
\vrd{D}_{ijkl} =\begin{pmatrix}
 1 & 0 & 0\\
 0 &-1 & 0\\
 0 & 0 &-1\\
\end{pmatrix}
\vrd{D}_{\sigma(i)\sigma(j)\sigma(k)\sigma(l)},\label{eq:mirror}
\end{equation}
where $\sigma(i)$ was defined in the context of Eq.~\eqref{eqn:mirroring}. Equation~(\ref{eq:mirror}) is satisfied for two-, three-, and four-site chiral interactions.

If Eq.~(\ref{eq:enerchiral}) is evaluated based on the model Eq.~\eqref{eq:Heis4}, one obtains
\begin{align}
\mathcal{D}^{\alpha}_{\text{r},ij} =& D_{ij}^\alpha + \sum_{k\notin \lbrace i,j \rbrace} D_{ikkj}^\alpha\nonumber\\
&+\sum_{k\neq l\notin \lbrace i,j \rbrace} \left( \frac{1}{2}D_{klij}^\alpha+D_{iklj}^\alpha\right).\label{eq:rotfsDM}
\end{align}
Note that the two-site four-spin interaction $\vrd{D}_{ijij}$ does not contribute to the grand potential of the configurations shown in Fig.~\ref{fig:initstates}, since it is only finite between pairs of spins which are not parallel and not perpendicular to each other. Considering the chiral interaction vectors in Table~\ref{tab:4spin} 
it is possible to derive
\begin{align}
\mathcal{D}^{x}_{\text{r},79}=\sum_{k\notin \lbrace 7,9 \rbrace} D_{7kk9}^{x}+\sum_{k\neq l\notin \lbrace 7,9 \rbrace} D_{7kl9}^{x} ,
\end{align}
which generally does not vanish since the three-site and four-site chiral interactions are not antisymmetric with respect to their first and fourth indices. This indicates the presence of four-spin chiral interactions in the system on the order of $0.4\,\text{meV}$, which is not negligible compared to the total value of chiral interactions also containing two-site contributions displayed in Table~\ref{tab:4spin}. The relation $\left(\mathcal{D}_{\text{r},78}^{x},\mathcal{D}_{\text{r},78}^{y},\mathcal{D}_{\text{r},78}^{z}\right)\neq\left(-\mathcal{D}_{\text{r},89}^{x},\mathcal{D}_{\text{r},89}^{y},\mathcal{D}_{\text{r},89}^{z}\right)$, which breaks the symmetry rules for two-spin DM interactions, may similarly be explained by the presence of the four-spin chiral interactions.

Changing from the two-spin model in Eq.~(\ref{eq:heis}) to the model containing also four-spin interactions in Eq.~(\ref{eq:Heis4}) naturally  modifies the interpretation of the chiral interaction energies obtained from the torque method,
\begin{align}
\mathcal{D}_{\text{t},ij}^{\alpha}=&D_{ij}^{\alpha}  +D_{ijij}^{\alpha} +\sum_{k\notin \lbrace i,j \rbrace} D_{kiij}^{\alpha}\nonumber \\
&-\sum_{k\notin \lbrace i,j \rbrace}D_{kjji}^{\alpha}  +\frac{1}{2}\sum_{k\neq l \notin \lbrace i,j \rbrace}D_{klij}^{\alpha} \, ;\label{eq:torquefsDM}
\end{align}
for details of the derivation see Appendix~\ref{sec:torque}. It should be noted that contrary to Eq.~(\ref{eq:rotfsDM}), Eq.~(\ref{eq:torquefsDM}) is antisymmetric with respect to the site indices $i$ and $j$, thus preserving the symmetries of the two-site DM vectors. Most importantly, the different treatment of the two-, three-, and four-site four-spin interactions between the rotational scheme in Eq.~(\ref{eq:rotfsDM}) and the torque method in Eq.~(\ref{eq:torquefsDM}) can explain why the $z$ components of the two kinds of chiral interaction vectors have different signs in Table~\ref{tab:4spin}. While the torque method supports positive $\chi_{i}^{z}$ or right-handed chirality in the $xy$ plane, the rotational scheme supports negative $\chi_{i}^{z}$ or left-handed chirality in the same plane. The chirality derived from the rotational method is in agreement with the lowest-energy spin spiral state obtained from MFT calculations and the ground state found by the \textit{ab initio} energy minimization discussed in Sec.~\ref{sec3b}.

\section{Conclusion\label{sec4}}

\textit{Ab initio} electronic structure calculations were performed to study the magnetic properties of Fe monatomic chains on the Re(0001) substrate. For all the considered chains strong antiferromagnetic couplings between the nearest-neighbor spins were observed, 
which in the case of the five-atom-long chain led to a nearly collinear antiferromagnetic ground state with the spins approximately aligned perpendicular to the chain and to the surface normal. 
As the length of the chain is increased, the frustration of the isotropic exchange interactions at farther neighbors transforms the ground state into a spin spiral state, with a single modulation period observable in the 15-atom-long chain. The experimental investigations in Ref.~\cite{Kim2018} also concluded on a spin spiral ground state of 40-atom-long Fe chains, although with a smaller period. 

For the 15-atom-long chain the spin components in the $xz$ plane were found to 
follow a right-handed rotation in the spin spiral. Regarding the rotation in the $xy$ plane, calculations based on a spin model only containing two-spin interactions yielded a 
right-handed 
rotation, while determining the optimal homogeneous planar spin spiral state using magnetic force theorem calculations and performing an optimization of the configuration directly within the \textit{ab initio} scheme both indicated a left-handed 
chirality. This discrepancy was resolved by considering chiral multispin interactions in the spin model, 
the presence of which in the system is supported by  calculations of specific rotated spin configurations sensitive to the chirality.

Since the experiments were carried out in the superconducting phase of the Re substrate,  future \textit{ab initio} calculations including the superconducting state of Re by solving the Bogoliubov--de~Gennes equation \cite{Csire2015,Csire2018}   
may reveal the reasons behind the observed discrepancies between theory and experiment regarding the spin spiral period. Such first-principles calculations would also enable the investigation of the interplay between exotic magnetic states and topological superconductivity in clusters of magnetic atoms on a superconducting substrate.

\begin{acknowledgments}

The authors would like to thank R. Wiesendanger for useful discussions. Financial support of the Nemzeti Kutat\'asi Fejleszt\'esi \'es Innov\'aci\'os Hivatal (Hungary) under Projects No. K115575 and No. FK124100, the BME-Nanotechnology FIKP grant of Emberi Er\H{o}forr\'asok Miniszt\'eriuma (BME FIKP-NAT), the Alexander von Humboldt Foundation, and the Deutsche Forschungsgemeinschaft via SFB668 are gratefully acknowledged.

\end{acknowledgments}

\appendix

\section{Torque method}

\label{sec:torque}

\begin{figure}[htb]
\begin{center}
\includegraphics[scale=1]{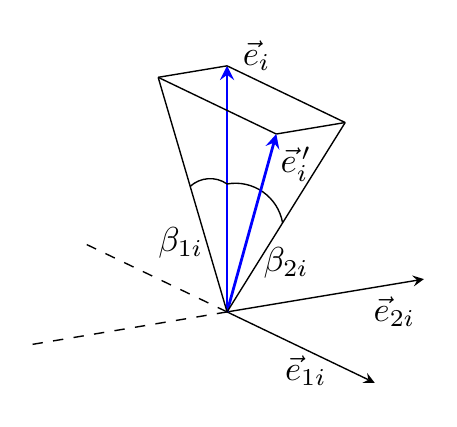}
\end{center}
\caption{Illustration of the rotation of the spin direction $\vrd{e}_i$ around the perpendicular vectors $\vrd{e}_{1i}$ and $\vrd{e}_{2i}$ by angles $\beta_{1i}$ and $\beta_{2i}$, used in the torque method to calculate derivatives of the grand potential. }
\label{fig:torque_vec}
\end{figure}

Here we give a short summary of how the parameters of the classical spin model in Eq.~\eqref{eq:heis} were determined by using the torque method; for more details see Ref.~\cite{Rozsa2016phd}. As shown in Fig.~\ref{fig:torque_vec}, at each site $i$ the spin direction will be denoted by $\vrd{e}_i$, and two orthogonal vectors $\vrd{e}_{1i}$, $\vrd{e}_{2i}$ are defined which form a right-handed basis together with $\vrd{e}_i$, and around which $\vrd{e}_i$ is rotated by the infinitesimal angles $\beta_{1i}$ and $\beta_{2i}$, respectively. The derivatives with respect to these angles may be expressed as
\begin{eqnarray}
\frac{\partial\:\:\:\:\:}{\partial \beta_{1i}}=-\vrd{e}_{2i}\frac{\partial\:\:\:}{\partial \vrd{e}_{i}}+\vrd{e}_{i}\frac{\partial\:\:\:\:\:}{\partial \vrd{e}_{2i}},\label{eqnS0a}
\\
\frac{\partial\:\:\:\:\:}{\partial \beta_{2i}}=\vrd{e}_{1i}\frac{\partial\:\:\:}{\partial \vrd{e}_{i}}-\vrd{e}_{i}\frac{\partial\:\:\:\:\:}{\partial \vrd{e}_{1i}}.\label{eqnS0b}
\end{eqnarray}

Supposing the model of Eq.~\eqref{eq:heis},
the second derivatives of $\Omega$ with respect to the angles $\beta_{1i}$ and $\beta_{2i}$ can be calculated as
\begin{equation}
\frac{\partial ^2 \Omega}{\partial \beta_{2j}\partial \beta_{2i}} = -J_{ij}^{\alpha\beta} e_{1i}^\alpha e_{1j}^\beta \qquad \text{for}~ j \neq i,\label{eqnS1}
\end{equation}
\begin{equation}
\frac{\partial ^2 \Omega}{\partial \beta_{2j}\partial \beta_{1i}} = J_{ij}^{\alpha\beta} e_{2i}^\alpha e_{1j}^\beta \qquad \text{for}~ j \neq i,\label{eqnS2}
\end{equation}
\begin{equation}
\frac{\partial ^2 \Omega}{\partial \beta_{1j}\partial \beta_{2i}} = J_{ij}^{\alpha\beta} e_{1i}^\alpha e_{2j}^\beta \qquad \text{for}~ j \neq i,\label{eqnS3}
\end{equation}
\begin{equation}
\frac{\partial ^2 \Omega}{\partial \beta_{1j}\partial \beta_{1i}} = -J_{ij}^{\alpha\beta} e_{2i}^\alpha e_{2j}^\beta \qquad \text{for}~ j \neq i,\label{eqnS4}
\end{equation}
\begin{equation}
\frac{\partial^2 \Omega}{\partial \beta_{2i}^2} =  \sum_j J_{ij}^{\alpha\beta} e_i^\alpha e_j^\beta - 2K_i^{\alpha\beta} e_i^\alpha e_i^\beta + 2K_i^{\alpha\beta} e_{1i}^\alpha e_{1i}^\beta ,\label{eqnS5}
\end{equation}
\begin{equation}
\frac{\partial^2 \Omega}{\partial \beta_{1i}^2} =  \sum_j J_{ij}^{\alpha\beta} e_i^\alpha e_j^\beta - 2K_i^{\alpha\beta} e_i^\alpha e_i^\beta + 2K_i^{\alpha\beta} e_{2i}^\alpha e_{2i}^\beta ,\label{eqnS6}
\end{equation}
\begin{equation}
\frac{\partial^2 \Omega}{\partial \beta_{1i}\partial\beta_{2i}} = -K_i^{\alpha\beta} \left( e_{1i}^\alpha e_{2i}^\beta + e_{2i}^\alpha e_{1i}^\beta \right),\label{eqnS7}
\end{equation}
where $\alpha$ and $\beta$ label summations over Cartesian indices.

Within the torque method the derivatives on the left-hand sides of Eqs.~(\ref{eqnS1})--(\ref{eqnS7}) are obtained directly from first principles, using the expressions derived from Lloyd's formula given in Ref.~\cite{Udvardi2003}. The diagonal elements of the anisotropy tensor $\mxrd{K}_i$ were determined from Eqs.~(\ref{eqnS5}) and (\ref{eqnS6}) via
\begin{eqnarray}
\frac{\partial^2 \Omega}{\partial \beta_{2i}^2}-\frac{\partial^2 \Omega}{\partial \beta_{1i}^2}=2K_i^{\alpha\beta} e_{1i}^\alpha e_{1i}^\beta-2K_i^{\alpha\beta} e_{2i}^\alpha e_{2i}^\beta,\label{eqnS7a}
\end{eqnarray}
which simplifies to $K_{i}^{yy}-K_{i}^{zz}$ if $\vrd{e}_i$ is pointing along the $x$ direction. This way it is only possible to determine the differences between the diagonal elements, but this information is sufficient since $\mxrd{K}_i$ was defined as a traceless tensor because its trace only shifts the grand potential by a constant factor due to the normalization of the spins. The two independent components were computed by considering spin configurations where all spins are pointing along the $x$ and $y$ directions.

For calculating the exchange interaction tensor $\mxrd{J}_{ij}$ one has to rely on Eqs.~(\ref{eqnS1})--(\ref{eqnS4}). By introducing the tensor products
\begin{eqnarray}
\vec{v}_{ri,pj}=\left(-1\right)^{r+p}\vec{e}_{ri}\otimes \vec{e}_{pj}\label{eqnS8}
\end{eqnarray}
for $r,p=1,2$, and the notation for the derivatives
\begin{eqnarray}
x_{ri,pj}=\frac{\partial ^2 \Omega}{\partial \beta_{\overline{p}j}\partial \beta_{\overline{r}i}},\label{eqnS9}
\end{eqnarray}
with $\overline{r}=3-r$ the opposite angle index, Eqs.~(\ref{eqnS1})--(\ref{eqnS4}) can be summarized as
\begin{eqnarray}
x_{ri,pj}=\vec{v}_{ri,pj}\vec{J}_{ij},\label{eqnS10}
\end{eqnarray}
where $\mxrd{J}_{ij}$ is rewritten as a vector $\vec{J}_{ij}$ in the tensor product space. By introducing $\mxrd{V}_{ri,pj}=\vec{v}_{ri,pj}\circ\vec{v}_{ri,pj}$, the orthogonal projection onto the subspace of $\vec{v}_{ri,pj}$, summing over Eqs.~(\ref{eqnS1})--(\ref{eqnS4}) one obtains the system of linear equations
\begin{eqnarray}
\sum_{r,p}x_{ri,pj}\vec{v}_{ri,pj}=\sum_{r,p}\mxrd{V}_{ri,pj}\vec{J}_{ij}.\label{eqnS11}
\end{eqnarray}
Since $\mxrd{V}_{ri,pj}$ is a rank-1 matrix, the sum over four equations in Eq.~(\ref{eqnS11}) only enables the calculation of four components of $\vec{J}_{ij}$ as described in Ref.~\cite{Udvardi2003}. In order to determine the full tensor, Eq.~(\ref{eqnS11}) has to be summed up over calculations performed for at least three linearly independent directions of the $\vrd{e}_i$ vectors, corresponding to a least-squares fitting procedure for $\vec{J}_{ij}$. Due to the $C_{3\textrm{v}}$  symmetry of the Re(0001) surface, during the calculations four ferromagnetic configurations of the $\vrd{e}_i$ vectors were taken into account, including three NN directions parallel to the $x$ axis and at angles $60^{\circ}$ and $120^{\circ}$ with respect to this direction, as well as one along the out-of-plane $z$ direction.


The chiral interaction vectors $\vrd{\mathcal{D}}_{\text{t},ij}$ introduced in Sec.~\ref{sec3c} were calculated in ferromagnetic configurations with all spins pointing along one of the $\alpha=x$, $y$, or $z$ directions, i.e. $\vrd{e}_{i}=\vrd{e}_{\alpha}$ in Fig~\ref{fig:torque_vec}. The components of $\vrd{\mathcal{D}}_{\text{t},ij}$ are defined as
\begin{equation}
\mathcal{D}_{\text{t},ij}^{\alpha} \equiv
\frac{1}{2}\left(\frac{\partial ^2 \Omega}{\partial \beta_{1j}\partial \beta_{2i}}-\frac{\partial ^2 \Omega}{\partial \beta_{2j}\partial \beta_{1i}}\right) \, , \label{eqnS12}
\end{equation}
which within the model Eq.~(\ref{eq:heis}) containing only two-spin interactions simplifies to the DM vector [cf. Eqs.~(\ref{eqnS2}) and (\ref{eqnS3})],
\begin{equation}
\mathcal{D}_{\text{t},ij}^{\alpha}=D_{ij}^{\alpha} \, . 
\end{equation}

On the other hand, inserting the spin model Eq.~(\ref{eq:Heis4}) instead of Eq.~(\ref{eq:heis}) into Eq.~\eqref{eqnS12} and using the relations Eqs.~(\ref{eqnS0a}) and (\ref{eqnS0b}) to calculate the second derivatives yields Eq.~(\ref{eq:torquefsDM}) in Sec.~\ref{sec3c},
\begin{align}
\mathcal{D}_{\text{t},ij}^{\alpha} =&D_{ij}^{\alpha}+D_{ijij}^{\alpha}+\sum_{k}D_{kiij}^{\alpha}\nonumber\\
&-\sum_{k}D_{kjji}^{\alpha}+\frac{1}{2}\sum_{k,l}D_{klij}^{\alpha} \,. \label{eqnS13}
\end{align}

\section{Alternative notation for the four-spin chiral interactions}

\label{sec:fourspin}

Equation~(\ref{eq:Heis4}) follows the convention where the summations are performed for each index separately over all lattice sites, treating the cases where some of the indices coincide individually. This is in agreement, e.g., with the notation for the two-site two-spin and four-spin chiral interactions used in Ref.~\cite{Brinker}. Another way of expressing the grand potential or the Hamiltonian is by performing the summations over all different plaquettes, i.e. pairs, triangles, and quadrilaterals in the case of the two-, three-, and four-site interactions, respectively. Such a convention is used, e.g., in Ref.~\cite{Hoffmann} for the isotropic four-spin interactions.

Following this convention, Eq.~(\ref{eq:Heis4}) may be rewritten as
\begin{eqnarray}
\Omega=&&\sum_{i} \vrd{e}_i \mxrd{K}_{i} \vrd{e}_i - \sum_{\left<i,j\right>} \vrd{e}_i \mxrd{J}_{ij} \vrd{e}_j-\sum_{\left<i,j\right>}\vrd{D}_{ijij}\left( \vrd{e}_{i} \vrd{e}_{j} \right)\left( \vrd{e}_{i} \times \vrd{e}_{j} \right)\nonumber
\\
-&&\sum_{\left<i,j,k\right>}\left[\vrd{D}_{ijjk}\left(\vrd{e}_{i}\vrd{e}_{j}\right)\left(\vrd{e}_{j}\times\vrd{e}_{k}\right)+\vrd{D}_{ikkj}\left(\vrd{e}_{i}\vrd{e}_{k}\right)\left(\vrd{e}_{k}\times\vrd{e}_{j}\right)\right.\nonumber
\\
&&\qquad \left.+\vrd{D}_{kjji}\left(\vrd{e}_{k}\vrd{e}_{j}\right)\left(\vrd{e}_{j}\times\vrd{e}_{i}\right)+\vrd{D}_{jkki}\left(\vrd{e}_{j}\vrd{e}_{k}\right)\left(\vrd{e}_{k}\times\vrd{e}_{i}\right)\right.\nonumber
\\ \nonumber \\[-5pt]
&& \qquad \left.+\vrd{D}_{jiik}\left(\vrd{e}_{j}\vrd{e}_{i}\right)\left(\vrd{e}_{i}\times\vrd{e}_{k}\right)+\vrd{D}_{kiij}\left(\vrd{e}_{k}\vrd{e}_{i}\right)\left(\vrd{e}_{i}\times\vrd{e}_{j}\right)\right]\nonumber
\\
-&&\sum_{\left<i,j,k,l\right>}\left[\vrd{D}_{ijkl}\left(\vrd{e}_{i}\vrd{e}_{j}\right)\left(\vrd{e}_{k}\times\vrd{e}_{l}\right)+\vrd{D}_{iklj}\left(\vrd{e}_{i}\vrd{e}_{k}\right)\left(\vrd{e}_{l}\times\vrd{e}_{j}\right)\right.\nonumber
\\ 
&& \qquad \; \left.+\vrd{D}_{ilkj}\left(\vrd{e}_{i}\vrd{e}_{l}\right)\left(\vrd{e}_{k}\times\vrd{e}_{j}\right)+\vrd{D}_{klij}\left(\vrd{e}_{k}\vrd{e}_{l}\right)\left(\vrd{e}_{i}\times\vrd{e}_{j}\right)\right.\nonumber
\\  \nonumber \\[-5pt]
&& \qquad \; \left.+\vrd{D}_{ljik}\left(\vrd{e}_{l}\vrd{e}_{j}\right)\left(\vrd{e}_{i}\times\vrd{e}_{k}\right)+\vrd{D}_{jkil}\left(\vrd{e}_{j}\vrd{e}_{k}\right)\left(\vrd{e}_{i}\times\vrd{e}_{l}\right)\right] .\nonumber
\\ \label{eqn60}
\end{eqnarray}

Once again, the $i$, $j$, $k$, and $l$ indices all label different atoms. The sums over pairs contain half as many terms as the two-site summations in Eq.~(\ref{eq:Heis4}); however, this is resolved by taking into account the intrinsic symmetry relations $\mxrd{J}_{ij}=\mxrd{J}_{ji}^T$ and Eq.~(\ref{twosite}), which enable canceling the $1/2$ prefactor. As discussed in the main text, no such intrinsic symmetry relations exist for the three-site four-spin chiral interaction, which necessitates a summation over six different terms in Eq.~(\ref{eqn60}), corresponding to the possible permutations of the site indices in a triangle. For the four-site interactions, the symmetry relations in Eq.~(\ref{foursite}) cancel with the prefactor $1/4$ in Eq.~(\ref{eq:Heis4}), simplifying the sum over $24$ permutations in a quadrilateral to six different terms in Eq.~(\ref{eqn60}).

Note that Eq.~(\ref{eqn60}) is the most general form of the grand potential containing four-spin chiral interactions. Crystal symmetries may reduce the number of independent coefficients for specific plaquettes. For example, the mirror symmetry defined in Eq.~(\ref{eq:mirror}) implies that in the triangle with $i=\sigma(j)$ and $k=\sigma(k)$ there are only three inequivalent three-site four-spin chiral interactions instead of six in the general case.

\section{Tilted spin spirals}

\label{sec:tilted}

In this section we summarize why the spin spiral ground state becomes tilted away from the $xy$ plane due to the $y$ component of the DM vectors. We will consider a simplified spin model consisting of $N$ sites where the interactions are homogeneous along the chain, with $K^{yy}<0$ anisotropy accounting for the easy $y$ direction and $D^{y}$ DM interaction between NN spins. We will compare the energies of harmonic spin spiral configurations as defined in Eq.~\eqref{eq:spiralstate}, simplifying the description to two parameters $\delta$ and $\alpha$. It is assumed that a spin spiral state with a specific $\delta$ value is formed by the frustration of the isotropic exchange interactions, as was discussed in Sec.~\ref{sec3b}, and only the dependence of the grand potential of the spin model on the $\alpha$ parameter is considered. This is given by the expression

\begin{align}
& \Omega\left(\alpha\right) = \Omega_{0} + \sum_{i=1}^N  K^{yy} \left( \cos\alpha \sin\left[ i \left( \pi\!+\!\delta \right) \!+\!\varphi_0 \right] \right) ^2 
 \nonumber \\
     & - \!\sum_{i=1}^{N-1}\!D^{y} \!\left(  \sin\alpha\sin \left[ i \left( \pi\!+\!\delta \right) \!+\! \varphi_0 \right] \cos\left[\left(i\!+\!1\right) \left( \pi\!+\!\delta \right) \!+\! \varphi_0 \right] \right. 
 \nonumber \\
     & \left. - \cos \left[ i \left( \pi\!+\!\delta \right) \!+\! \varphi_0 \right] \sin\alpha\sin\left[\left(i\!+\!1\right) \left( \pi\!+\!\delta \right)\!+\! \varphi_0 \right] \right)  ,
\end{align}

\noindent
where now $\Omega_{0}$ describes the contributions which do not depend on $\alpha$, such as the isotropic interactions. For $N \geq 3 $, the average of $\sin^2\left[ i\left( \pi+\delta \right) +\varphi_0 \right] $ with respect to atomic indices equals $1/2$, and using a simple addition formula it is possible to arrive at

\begin{equation}
\Omega\left(\alpha\right) = \Omega_{0} + \frac{1}{2}NK^{yy} \cos^2\alpha - \left( N-1 \right) D^{y} \sin\alpha \sin \delta .
\end{equation}

One can obtain the value of $\alpha$ minimizing the grand potential by differentiation,

\begin{equation}
 \alpha = \arcsin \left( -\frac{N-1}{N}\frac{\sin\delta D^{y}}{K^{yy}} \right)
\end{equation}

\noindent
for $\left|\left( N-1 \right)\sin\delta D^{y}/NK^{yy}\right|\le 1$, and

\begin{equation}
 \alpha = \pm \frac{\pi}{2}
\end{equation}

\noindent
otherwise. This indicates that for an arbitrarily small value of $D^{y}$, the most preferred state is tilted away from the $xy$ plane where $\alpha=0$. We note that including the $z$ component of the DM vectors influences the dependence of the grand potential on the $\alpha$ parameter, but does not change this qualitative conclusion.

\bibliographystyle{apsrev4-1}
\bibliography{fere}

\end{document}